\documentclass[aps,prd,twocolumn,superscriptaddress,nofootinbib]{revtex4-1}

\usepackage{latexsym}
\usepackage{amsmath}
\usepackage{amssymb}
\usepackage{amsfonts}
\usepackage{bm}

\usepackage{color}
\definecolor{purple}{rgb}{0.5,0,0.5}
\definecolor{blue}{rgb}{0.0,0,0.9}
\definecolor{prdblue}{rgb}{0.133,0.118,0.498}
\usepackage[colorlinks=true, pdfstartview=FitV, linkcolor=prdblue, citecolor= prdblue, urlcolor=prdblue]{hyperref}

\usepackage{supertabular} 
\usepackage{placeins}
\usepackage{epsfig}
\usepackage{graphicx}

\begin{document}


\title{Kaon spectrum revisited}


\author{U. Taboada-Nieto}
\email[]{id00746851@usal.es}
\affiliation{Departamento de Física Fundamental, Universidad de Salamanca, E-37008 Salamanca, Spain}

\author{P. G. Ortega}
\email[]{pgortega@usal.es}
\affiliation{Departamento de Física Fundamental, Universidad de Salamanca, E-37008 Salamanca, Spain}
\affiliation{Instituto Universitario de F\'isica Fundamental y Matem\'aticas (IUFFyM), Universidad de Salamanca, E-37008 Salamanca, Spain}

\author{D. R. Entem}
\email[]{entem@usal.es}
\affiliation{Departamento de Física Fundamental, Universidad de Salamanca, E-37008 Salamanca, Spain}
\affiliation{Instituto Universitario de F\'isica Fundamental y Matem\'aticas (IUFFyM), Universidad de Salamanca, E-37008 Salamanca, Spain}

\author{F. Fern\'andez}
\email[]{fdz@usal.es}
\affiliation{Departamento de Física Fundamental, Universidad de Salamanca, E-37008 Salamanca, Spain}
\affiliation{Instituto Universitario de F\'isica Fundamental y Matem\'aticas (IUFFyM), Universidad de Salamanca, E-37008 Salamanca, Spain}

\author{J. Segovia}
\email[]{jsegovia@upo.es}
\affiliation{Departamento de Sistemas F\'isicos, Qu\'imicos y Naturales, \\ Universidad Pablo de Olavide, E-41013 Sevilla, Spain}


\date{\today}

\begin{abstract}
The European Organization for Nuclear Research (CERN) has recently approved a world-unique QCD facility in which an updated version of the external M2 beam line of the CERN SPS in conjunction with a universal spectrometer of the COMPASS experiment is used. One of its main goals is to use highly intense and energetic kaon beams to map out the complete spectrum of excited kaons with an unprecedented precision; having a broad impact not only on low-energy QCD phenomenology, but also on many high-energy particle processes where excited kaons appear, such as the study of CP violation in heavy-meson decays studied at LHCb and Belle~II.
In support of the experimental effort, the kaon spectrum is computed herein using a constituent quark model which has been successfully applied to a wide range of hadronic observables, from light to heavy quark sectors, and thus the model parameters are completely constrained.
The model's prediction can be used as a template against which to compare the already collected data and future experimental findings, in order to distinguish between conventional and exotic kaon states. We also compare our results with those available in the literature in order to provide some general statements, common to all calculations.
\end{abstract}

\keywords{
Quantum Chromodynamics \and
Quark model            \and
}

\maketitle


\section{Introduction}
\label{sec:intro}

Most of the data on strange mesons are based on experiments that were performed around 30 years ago. Nevertheless, strange mesons appear in many high-energy particle processes as searches for CP violation in multi-body heavy-meson decays like $B^\pm \to D^0 K^\pm \to (K_S^0\pi^+\pi^-)K^\pm$ performed by the BaBar~\cite{BaBar:2008inr}, Belle~\cite{Belle:2010xyn} and LHCb~\cite{LHCb:2014kxv} collaborations. Nowadays, the Particle Data Group (PDG) lists $25$ strange mesons that have been detected within the energy region [0.5,3.1]~GeV~\cite{ParticleDataGroup:2020ssz}. Among the $25$ states, only $14$ are well established and included in the summary tables whereas the remaining $11$ states still need further confirmation; with two of them having their spin-parity quantum numbers not yet determined.

A completely new horizon in kaon physics has been opened with the CERN's approval of upgrading the external M2 beam line of the CERN SPS in order to provide radio-frequency separated high-intensity and high-energy kaon and antiproton beams. The kaon beam, in combination with a universal spectrometer of the COMPASS experiment, will allow to map out the complete spectrum of excited kaons with unprecedented precision, using novel analysis methods. 

There are also proposals and plans for future measurements of strange mesons using either $\tau$ or $D$-meson decays at other facilities such as Belle~II, BES~III and LHCb. These studies will be limited by the mass of the initial state which is around 2~GeV in both cases. Besides, a new kaon beam line will be built at J-PARC in the near future~\cite{Aoki:2021cqa}; however, the expected low momenta would make it difficult to separate beam and target excitations, which might lead to large systematic uncertainties in the determination of strange-meson spectrum.

In support of the experimental effort, we revisit herein the kaon spectrum using a constituent quark model~\cite{Vijande:2004he, Segovia:2013wma} in which the quark-antiquark interaction is based on the non-perturbative phenomena of dynamical chiral symmetry breaking and color confinement, plus the perturbative one-gluon exchange (OGE) force. As it is well known, the quark model parameters are crucial; we have used values that have been fitted before through hadron~\cite{Valcarce:1995dm, Valcarce:2005rr, Segovia:2011zza, Segovia:2015dia, Yang:2019lsg}, hadron-hadron~\cite{Entem:2000mq, Valcarce:2005em, Ortega:2016pgg, Ortega:2018cnm} and multiquark~\cite{Vijande:2003ki, Yang:2018oqd, Yang:2020twg, Ortega:2021xst, Yang:2021zhe} phenomenology, from light to heavy quark sectors, and thus the model parameters are completely constrained. The model can be used as a template against which to compare the already collected data of kaons in the PDG, and the future experimental findings, in order to distinguish between conventional quark-antiquark bound states and those whose nature could be in conflict with standard quarkonium interpretations.

The bound-state problem is solved using the Rayleigh-Ritz variational method in which the genuine state's wave function is expanded by means of Gaussian trial functions whose ranges are in geometric progression~\cite{Hiyama:2003cu}. This provides enough accuracy and it makes the subsequent evaluation of the decay amplitude matrix elements easier. Moreover, the geometric progression is dense at short distances, so that it allows the description of the dynamics mediated by short range potentials. The fast damping of the gaussian tail is not a problem, since we can choose the maximal range much longer than the hadronic size.

The manuscript is arranged as follows. After this introduction, the theoretical framework is presented in Sec.~\ref{sec:model}. Section~\ref{sec:results} is mostly devoted to the analysis and discussion of our theoretical results; we end this section by comparing our spectrum with those produced by other theoretical frameworks. Finally, we summarize and give some prospects in Sec.~\ref{sec:summary}.


\section{Theoretical framework}
\label{sec:model} 

We describe in this section the basic phenomenological properties of QCD and how they are implemented into the constituent quark model. There is a detailed description of all different terms of the interacting potential and the variational method used to solve the Schr\"odinger equation. This section also contains a discussion related with the convenience of using a non-relativistic approach to the kaon sector.

\subsection{CONSTITUENT QUARK MODEL}
\label{sec:CQM}

A consequence of the spontaneous chiral symmetry breaking is that the nearly
massless current light quarks that appear in the QCD Lagrangian acquire a constituent quark mass, $M(p)$, which is momentum dependent. To preserve chiral invariance of the QCD Lagrangian new interaction terms appear between constituent quarks which are given by Goldstone boson exchanges.

A simplistic Lagrangian invariant under chiral transformations can be derived as~\cite{Diakonov:2002fq}
\begin{eqnarray}
{\mathcal L} &=& \bar \psi ( i\gamma^\mu \partial_\mu - M U^{\gamma_5})\psi \,,
\end{eqnarray}
where $U^{\gamma_5}=\exp(i\pi^a\lambda^a\gamma_5/f_\pi)$ with $\pi^a$ denoting the pseudoscalar fields $(\vec \pi,K,\eta_8)$ and $M$ the constituent quark mass. The momentum dependent constituent quark mass acts as a natural cut-off of the theory. The chiral quark-quark interaction can be written in the following way
\begin{align}
V_{qq}\left(\vec{r}_{ij}\right) = V_{qq}^{C}\left(\vec{r}_{ij}\right) + V_{qq}^{T}\left(\vec{r}_{ij}\right) + V_{qq}^{SO}\left(\vec{r}_{ij}\right) \,,
\end{align}
where $C$, $T$ and $SO$ stand for central, tensor and spin-orbit potentials. The central contribution has four different terms,
\begin{equation}
V_{qq}^{C}\left(\vec{r}_{ij}\right) = V_{\pi}^{C}\left(\vec{r}_{ij}\right) + V_{\sigma}^{C}\left(\vec{r}_{ij}\right) + V_{K}^{C}\left(\vec{r}_{ij}\right) + V_{\eta}^{C}\left(\vec{r}_{ij}\right),
\end{equation}
which are given by the following expressions
\begin{widetext}
\begin{subequations}
\begin{align}
&
V_{\pi}^{C}\left( \vec{r}_{ij} \right)= \frac{g_{ch}^{2}}{4\pi} \frac{m_{\pi}^2}{12m_{i}m_{j}}\frac{\Lambda_{\pi}^{2}}{\Lambda_{\pi}^{2}-m_{\pi}^{2}}m_{\pi}\left[Y(m_{\pi}r_{ij})-\frac{\Lambda_{\pi}^{3}}{m_{\pi}^{3}} Y(\Lambda_{\pi}r_{ij}) \right] \times (\vec{\sigma}_{i}\cdot\vec{\sigma}_{j})\sum_{a=1}^{3}(\lambda_{i}^{a} \cdot\lambda_{j}^{a}) \,, \\
& 
V_{\sigma}^{C}\left( \vec{r}_{ij} \right) = -\frac{g_{ch}^{2}}{4\pi} \frac{\Lambda_{\sigma}^{2}}{\Lambda_{\sigma}^{2}-m_{\sigma}^{2}} m_{\sigma} \left[ Y(m_{\sigma}r_{ij}) - \frac{\Lambda_{\sigma}}{m_{\sigma}}Y(\Lambda_{\sigma}r_{ij}) \right] \,, \\
& 
V_{K}^{C}\left( \vec{r}_{ij} \right)= \frac{g_{ch}^{2}}{4\pi} \frac{m_{K}^2}{12m_{i}m_{j}}\frac{\Lambda_{K}^{2}}{\Lambda_{K}^{2}-m_{K}^{2}}m_{K}\left[Y(m_{K}r_{ij})-\frac{\Lambda_{K}^{3}}{m_{K}^{3}}Y(\Lambda_{K}r_{ij}) \right] \times (\vec{\sigma}_{i}\cdot\vec{\sigma}_{j})\sum_{a=4}^{7}(\lambda_{i}^{a} \cdot\lambda_{j}^{a}) \,, \\
& 
V_{\eta}^{C}\left( \vec{r}_{ij} \right)= \frac{g_{ch}^{2}}{4\pi} \frac{m_{\eta}^2}{12m_{i}m_{j}}\frac{\Lambda_{\eta}^{2}}{\Lambda_{\eta}^{2}-m_{\eta}^{2}}m_{\eta}\left[Y(m_{\eta}r_{ij})-\frac{\Lambda_{\eta}^{3}}{m_{\eta}^{3}}Y(\Lambda_{\eta}r_{ij}) \right] \times (\vec{\sigma}_{i}\cdot\vec{\sigma}_{j}) \left[\cos\theta_{p}\left(\lambda_{i}^{8}\cdot\lambda_{j}^{8} \right)-\sin\theta_{p}\right] \,.
\end{align}
\end{subequations}
\end{widetext}
We consider the physical $\eta$-meson instead of the octet one and thus the angle $\theta_p$ is introduced. The standard Yukawa function is defined by $Y(x)=e^{-x}/x$. The $\lambda^{a}$ are the SU(3) flavour Gell-Mann matrices. The quark masses are $m_{i}$ and $m_{\pi}$, $m_{K}$ and $m_{\eta}$ are the masses of the SU(3) Goldstone bosons, taken from experimental values. The value of $m_\sigma$ is given by the partially conserved axial current relation $m_{\sigma}^{2}\simeq m_{\pi}^{2}+4m_{u,d}^{2}$~\cite{Scadron:1982eg}. Note, however, that better determinations of the mass of the $\sigma$-meson have been reported since then~\cite{Garcia-Martin:2011nna, Albaladejo:2012te} -- see also the recent review~\cite{Pelaez:2015qba}; one should simply consider the value used here as a model parameter. Finally, the chiral coupling
constant, $g_{ch}$, is determined from the $\pi NN$ coupling constant through
\begin{equation}
\frac{g_{ch}^{2}}{4\pi} = \frac{9}{25}\frac{g_{\pi NN}^{2}}{4\pi} \frac{m_{u,d}^{2}}{m_{N}^2},
\end{equation}
which assumes that flavour SU(3) is an exact symmetry only broken by the different mass of the strange quark.

There are three different contributions to the tensor potential:
\begin{equation}
V_{qq}^T(\vec{r}_{ij}) = V_{\pi}^{T}(\vec{r}_{ij}) + V_{K}^{T}(\vec{r}_{ij})
+ V_{\eta}^{T}(\vec{r}_{ij}),
\end{equation}
given by
\begin{widetext}
\begin{subequations}
\begin{align}
&
V_{\pi}^{T}\left( \vec{r}_{ij} \right)= \frac{g_{ch}^{2}}{4\pi} \frac{m_{\pi}^2}{12m_{i}m_{j}}\frac{\Lambda_{\pi}^{2}}{\Lambda_{\pi}^{2}-m_{\pi}^{2}}m_{\pi}\left[H(m_{\pi}r_{ij})-\frac{\Lambda_{\pi}^{3}}{m_{\pi}^{3}} H(\Lambda_{\pi}r_{ij}) \right] \times S_{ij} \sum_{a=1}^{3}(\lambda_{i}^{a}\cdot\lambda_{j}^{a}), \\
& 
V_{K}^{T}\left( \vec{r}_{ij} \right)= \frac{g_{ch}^{2}}{4\pi} \frac{m_{K}^2}{12m_{i}m_{j}}\frac{\Lambda_{K}^{2}}{\Lambda_{K}^{2}-m_{K}^{2}}m_{K}\left[H(m_{K}r_{ij})-\frac{\Lambda_{K}^{3}}{m_{K}^{3}}H(\Lambda_{K}r_{ij}) \right] \times S_{ij} \sum_{a=4}^{7}(\lambda_{i}^{a}\cdot\lambda_{j}^{a}), \\
& 
V_{\eta}^{T}\left( \vec{r}_{ij} \right)= \frac{g_{ch}^{2}}{4\pi} \frac{m_{\eta}^2}{12m_{i}m_{j}}\frac{\Lambda_{\eta}^{2}}{ \Lambda_{\eta}^{2}-m_{\eta}^{2}}m_{\eta}\left[H(m_{\eta}r_{ij})-\frac{\Lambda_{\eta}^{3}}{m_{\eta}^{3}}H(\Lambda_{\eta}r_{ij}) \right] \times S_{ij} \left[ \cos \theta_{p} \left(\lambda_{i}^{8}\cdot\lambda_{j}^{8} \right)-\sin\theta_{p}\right].
\end{align}
\end{subequations}
\end{widetext}
The quark tensor operator is $S_{ij}=3(\vec{\sigma}_{i}\cdot\hat{r}_{ij})(\vec{\sigma}_{j}\cdot\hat{r}_{ij}) - \vec{\sigma_{i}}\cdot\vec{\sigma_{j}}$ and $H(x)=(1+3/x+3/x^{2})Y(x)$.

Finally, the spin-orbit potential only presents a contribution coming from the scalar part of the interaction
\begin{align}
&
V_{qq}^{SO}(\vec{r}_{ij})=V_{\sigma}^{SO}\left( \vec{r}_{ij} \right) = - \frac{g_{ch}^{2}}{4\pi}\frac{m_{\sigma}^{3}}{2m_{i}m_{j}}\frac{\Lambda_{\sigma}^{2}}{\Lambda_{\sigma}^{2}-m_{\sigma}^{2}} \nonumber \\ 
&
\times\left[G(m_{\sigma}r_{ij})-\frac{\Lambda_{\sigma}^{3}}{m_{\sigma}^{3}} G(\Lambda_{\sigma}r_{ij}) \right] (\vec{L}\cdot\vec{S}).
\end{align}
In the last equation $G(x)$ is the function $(1+1/x)Y(x)/x$.

Beyond the scale of chiral symmetry breaking, the dynamics should be governed by QCD perturbative effects. They are taken into account by one-gluon fluctuations around the instanton vacuum. The effective vertex Lagrangian that define such effect is given by
\begin{eqnarray}
{\mathcal L}_{qqg} &=& i\sqrt{4\pi\alpha_s} \bar \psi \gamma_\mu G^\mu_c \lambda^c \psi,
\label{Lqqg}
\end{eqnarray}
with $\lambda^c$ being the $SU(3)$ color matrices and $G^\mu_c$ the gluon field.

The different terms of the potential derived from the Lagrangian above contain
central, tensor and spin-orbit contributions. They can be written in the following way
\begin{widetext}
\begin{subequations}
\begin{align}
V_{\rm OGE}^{\rm C}(\vec{r}_{ij}) = & \frac{1}{4}\alpha_{s}(\vec{\lambda}_{i}^{c}\cdot \vec{\lambda}_{j}^{c})\left[ \frac{1}{r_{ij}}-\frac{1}{6m_{i}m_{j}}  (\vec{\sigma}_{i}\cdot\vec{\sigma}_{j})  \frac{e^{-r_{ij}/r_{0}(\mu)}}{r_{ij}r_{0}^{2}(\mu)}\right], \\
V_{\rm OGE}^{\rm T}(\vec{r}_{ij})= &-\frac{1}{16}\frac{\alpha_{s}}{m_{i}m_{j}} (\vec{\lambda}_{i}^{c}\cdot\vec{\lambda}_{j}^{c})\left[  \frac{1}{r_{ij}^{3}}-\frac{e^{-r_{ij}/r_{g}(\mu)}}{r_{ij}}\left(  \frac{1}{r_{ij}^{2}}+\frac{1}{3r_{g}^{2}(\mu)}+\frac{1}{r_{ij}r_{g}(\mu)}\right) \right]S_{ij}, \\
V_{\rm OGE}^{\rm SO}(\vec{r}_{ij}) = & -\frac{1}{16}\frac{\alpha_{s}}{m_{i}^{2}m_{j}^{2}}(\vec{\lambda}_{i}^{c}\cdot \vec{\lambda}_{j}^{c})\left[\frac{1}{r_{ij}^{3}}-\frac{e^{-r_{ij}/r_{g}(\mu)}}{ r_{ij}^{3}} \left(1+\frac{r_{ij}}{r_{g}(\mu)}\right)\right] \times \nonumber \\
&
\times \left[((m_{i}+m_{j})^{2}+2m_{i}m_{j})(\vec{S}_{+}\cdot\vec{L})+(m_{j}^{2}
-m_{i}^{2}) (\vec{S}_{-}\cdot\vec{L}) \right],
\end{align}
\end{subequations}
\end{widetext}
where $\vec{S}_{\pm}=\frac{1}{2}(\vec{\sigma}_{i}\,\pm\,\vec{\sigma}_{j})$ are the so-called symmetric and anti-symmetric spin-orbit operators, respectively.
Besides, $r_{0}(\mu)=\hat{r}_{0}\frac{\mu_{nn}}{\mu_{ij}}$ and $r_{g}(\mu)=\hat{r}_{g}\frac{\mu_{nn}}{\mu_{ij}}$ are space-regulators which depend on $\mu_{ij}$, the reduced mass of the quark-antiquark pair. The contact term of the central potential has been regularized as
\begin{equation}
\delta(\vec{r}_{ij})\sim\frac{1}{4\pi r_{0}^{2}}\frac{e^{-r_{ij}/r_{0}}}{r_{ij}}
\end{equation}

The wide energy range needed to provide a consistent description of light,
strange and heavy mesons requires an effective scale-dependent strong coupling
constant. We use the frozen coupling constant of Ref.~\cite{Vijande:2004he}:
\begin{equation}
\alpha_{s}(\mu_{ij})=\frac{\alpha_{0}}{\ln\left( 
\frac{\mu_{ij}^{2}+\mu_{0}^{2}}{\Lambda_{0}^{2}} \right)},
\end{equation}
in $\alpha_{0}$, $\mu_{0}$ and $\Lambda_{0}$ are parameters of the model determined by a global fit to the meson spectra.

Color confinement is an empirical fact in which quarks, antiquarks and gluons appear always inside colorless structures named hadrons. It is well known that multi-gluon exchanges produce an attractive linearly rising potential proportional to the distance between infinite-heavy quarks. However, sea quarks are also important ingredients of the strong interaction dynamics that contribute to the screening of the rising potential at low momenta and eventually to the breaking of the quark-antiquark binding string~\cite{Bali:2005fu}. All these facts have been taken into account in our model by including the terms
\begin{widetext}
\begin{subequations}
\begin{align}
V_{\rm CON}^{\rm C}(\vec{r}_{ij}) =& \left[ -a_{c}(1-e^{-\mu_{c}r_{ij}})+\Delta
\right] (\vec{\lambda}_{i}^{c}\cdot\vec{\lambda}_{j}^{c}), \\
V_{\rm CON}^{\rm SO}(\vec{r}_{ij})= &
-\left(\vec{\lambda}_{i}^{c}\cdot\vec{\lambda}_{j}^{c} \right) 
\frac{a_{c}\mu_{c}e^{-\mu_{c}r_{ij}}}{4m_{i}^{2}m_{j}^{2}r_{ij}}\left[((m_{i}^{2
}+m_{j}^{2})(1-2a_{s}) +4m_{i}m_{j}(1-a_{s}))(\vec{S}_{+}\cdot\vec{L})\right.
\nonumber \\
&
\left. +(m_{j}^{2}-m_{i}^{2})(1-2a_{s})(\vec{S}_{-}\cdot\vec{L}) \right], 
\end{align}
\end{subequations}
\end{widetext}
where $a_{s}$ controls the mixture between the scalar and vector Lorentz
structures of the confinement. At short distances this potential presents a
linear behavior with an effective confinement strength
$\sigma=-a_{c}\,\mu_{c}\,(\vec{\lambda}^{c}_{i} \cdot \vec{\lambda}^{c}_{j})$
and becomes constant at large distances with a threshold defined by
\begin{equation}
V_{\rm thr}=\{-a_{c}+ \Delta\}(\vec{\lambda}^{c}_{i}\cdot
\vec{\lambda}^{c}_{j}).
\label{eq:threshold}
\end{equation}
No bound states can be found for energies higher than this threshold.
The system suffers a transition from a color string configuration between two
static color sources into a pair of static mesons due to the breaking of the
color string and the most favored decay into hadrons.

Among the different methods to solve the Schr\"odinger equation in order to 
find the quark-antiquark bound states, we use the Gaussian Expansion
Method~\cite{Hiyama:2003cu} because it provides enough accuracy and it makes the
subsequent evaluation of the decay amplitude matrix elements easier. 

This procedure provides the radial wave function solution of the Schr\"odinger
equation as an expansion in terms of basis functions
\begin{equation}
R_{\alpha}(r)=\sum_{n=1}^{n_{max}} c_{n}^\alpha \phi^G_{nl}(r),
\end{equation} 
where $\alpha$ refers to the channel quantum numbers. The coefficients,
$c_{n}^\alpha$, and the eigenvalue, $E$, are determined from the Rayleigh-Ritz
variational principle
\begin{equation}
\sum_{n=1}^{n_{max}} \left[\left(T_{n'n}^\alpha-EN_{n'n}^\alpha\right)
c_{n}^\alpha+\sum_{\alpha'}
\ V_{n'n}^{\alpha\alpha'}c_{n}^{\alpha'}=0\right],
\end{equation}
where $T_{n'n}^\alpha$, $N_{n'n}^\alpha$ and $V_{n'n}^{\alpha\alpha'}$ are the 
matrix elements of the kinetic energy, the normalization and the potential, 
respectively. $T_{n'n}^\alpha$ and $N_{n'n}^\alpha$ are diagonal whereas the
mixing between different channels is given by $V_{n'n}^{\alpha\alpha'}$.

\begin{table}[!t]
\caption{\label{tab:parameters} Quark model parameters.}
\begin{ruledtabular}
\begin{tabular}{lrr}
Quark masses    & $m_{n}$ (MeV) & $313$ \\
 		 & $m_{s}$ (MeV) & $555$ \\
\hline
Goldstone Bosons & $m_{\pi}$ $(\mbox{fm}^{-1})$ & $0.70$ \\
		  & $m_{\sigma}$ $(\mbox{fm}^{-1})$ & $3.42$ \\
 		  & $m_{K}$ $(\mbox{fm}^{-1})$ & $2.51$ \\
 		  & $m_{\eta}$ $(\mbox{fm}^{-1})$ & $2.77$ \\
 		  & $\Lambda_{\pi}$ $(\mbox{fm}^{-1})$ & $4.20$ \\
		  & $\Lambda_{\sigma}$ $(\mbox{fm}^{-1})$ & $4.20$ \\
 		  & $\Lambda_{K}$ $(\mbox{fm}^{-1})$ & $4.20$ \\
		  & $\Lambda_{\eta}$ $(\mbox{fm}^{-1})$ & $4.20$ \\
 		  & $g^{2}_{ch}/4\pi$ & $0.54$ \\
 		  & $\theta_{p}$ $(^\circ)$ & $-15$ \\
\hline
OGE & $\alpha_{0}$ & $2.118$ \\
    & $\Lambda_{0}$ $(\mbox{fm}^{-1})$ & $0.113$ \\
    & $\mu_{0}$ (MeV) & $36.976$ \\
    & $\hat{r}_{0}$ (fm) & $0.181$ \\
    & $\hat{r}_{g}$ (fm) & $0.259$ \\
\hline
Confinement & $a_{c}$ (MeV) & $507.4$ \\
            & $\mu_{c}$ $(\mbox{fm}^{-1})$ & $0.576$ \\
            & $\Delta$ (MeV) & $184.432$ \\
            & $a_{s}$ & $0.81$ \\
\end{tabular}
\end{ruledtabular}
\end{table}

Following Ref.~\cite{Hiyama:2003cu}, we employ Gaussian trial functions whose
ranges are in geometric progression. This enables the optimization of ranges
employing a small number of free parameters. Moreover, the geometric
progression is dense at short distances, so that it allows the description of
the dynamics mediated by short range potentials. The fast damping of the
gaussian tail is not a problem, since we can choose the maximal range much
longer than the hadronic size.

The model parameters fitted over all meson spectra and relevant for this work are shown in Table~\ref{tab:parameters}.


\subsection{Relativity and model-independence}
\label{sec:relativity}

Model estimates of the mean momentum, $\langle p \rangle$, of a light constituent quark, with mass $M$, inside a meson typically yield $\langle p \rangle \sim M$. It might therefore be argued that bound-state calculations involving light quark systems should only be undertaken within models that, at some level, incorporate relativity. This potential weakness of the nonrelativistic quark model has long been considered. For example, Ref.~\cite{Capstick:1985xss} remarks that a non-relativistic treatment of quark motion is inaccurate. However, using scales that are internally consistent, it is not ultra-relativistic. Therefore, the non-relativistic approximation must be useful. The point is also canvassed in Ref.~\cite{Manohar:1983md}, which opens with the question ``Why does the non-relativistic quark model work?" and proceeds to provide a range of plausible answers. These discussions are complemented by Ref.~\cite{Lucha:1991vn}, which devotes itself to ``The significance of the treatment of relativistically moving constituents by an effective non-relativistic Schr\"odinger equation [...]". The conclusion of these discourses and many others is simple: the non-relativistic model has proved very useful, unifying a wide range of observables within a single framework.

This last observation provides our rationale for employing a non-relativistic model for the analysis herein. Namely, we take a pragmatic view: the non-relativistic quark model is a useful tool. The practical reason for its
success is simple: the model has many parameters; they are fitted to a body of data; and, consequently, on this domain, the model cannot be wrong numerically. If one adds relativistic effects in one way or another, there are similar parameters in the new potential. They, too, are fitted to data; and hence the resulting model cannot produce results that are very different from the original non-relativistic version. The values of the parameters in the potential are modified, but the potential is not observable, so nothing substantive is altered.


\section{Results}
\label{sec:results}

Beyond the renewed experimental interest in studying the spectrum of kaons, its theoretical study is interesting for several reasons. For instance, the strange mesons exhibit explicit flavor, and thus $C$-parity cannot be a good quantum number, resulting in the fact that $J^{PC}$-exotic quark-gluon hybrid kaons should appear as an overpopulation of states in the conventional excited kaon spectrum. An example of this is the $J^P=1^-$ channel that should exhibit a rich number of kaon-like states, with some of them not fitting the quark model expectations, because the $J^{PC}=1^{-+}$ exotics are included beside the conventional $1^{--}$ states. Another important feature of kaons, also related with the fact that they exhibit explicit flavor, is the non-presence of quark-antiquark annihilation which makes its spectrum an ideal ground to compare with the isoscalar light-meson sector, where $q\bar q$ states can mix with the so-called glueballs.

\begin{figure*}[!t]
\includegraphics[width=\textwidth, height=0.66\textheight]{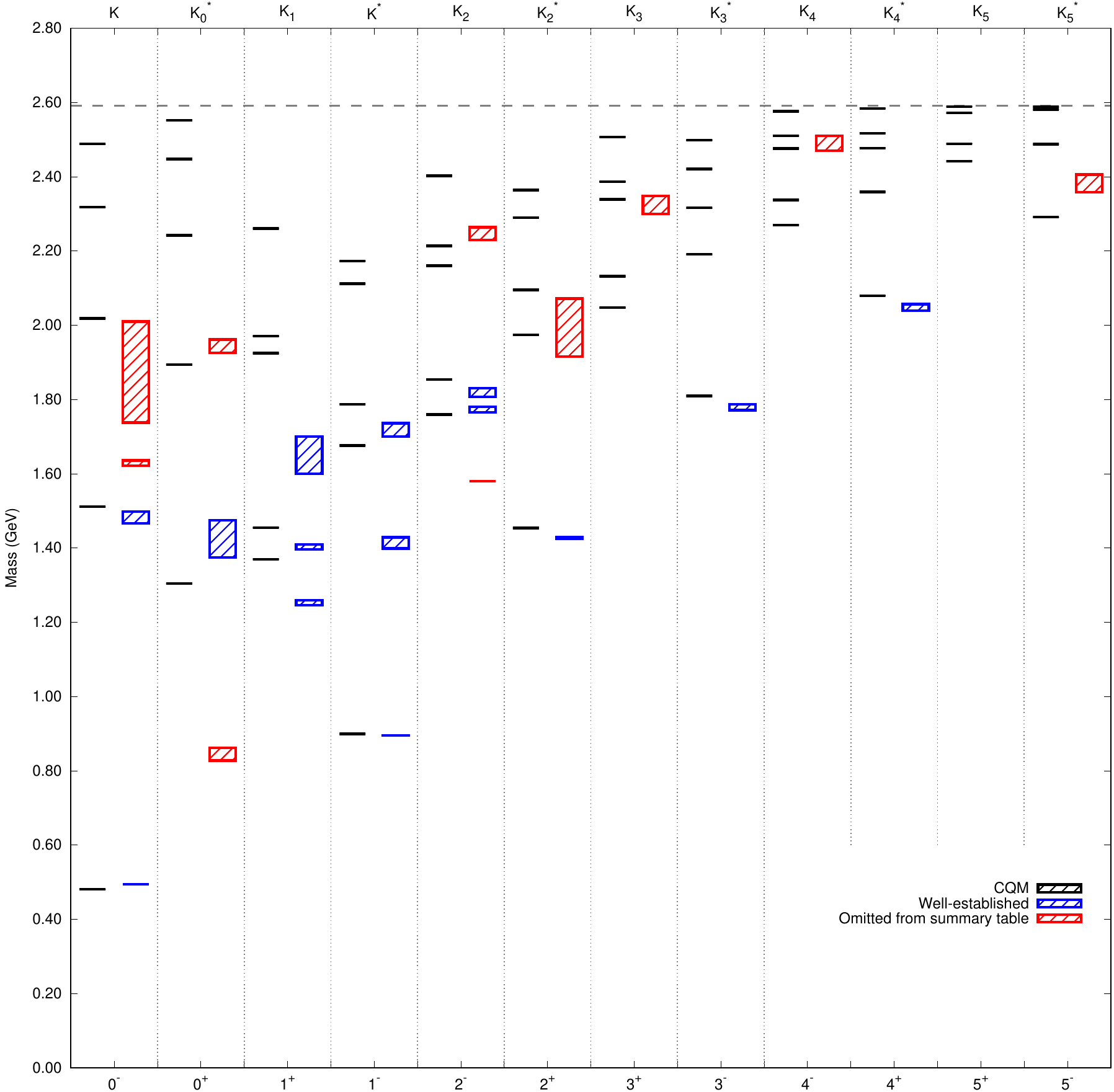}
\caption{\label{fig:spectrum} Spectrum of strange mesons predicted by our constituent quark model (black solid lines). It is compared with that collected in PDG~\cite{ParticleDataGroup:2020ssz}. States included in the PDG summary table are shown in blue, the remaining one are shown in red. The states are grouped by their spin-parity $J^P$ quantum numbers, and named accordingly. It is worth noting that our spectrum is finite because the energy of the quark-antiquark bound-state cannot be larger than $m_q + m_{\bar q} + V_{\text{thr}} = 2591\,\text{MeV}$, which is marked by a dashed line.}
\end{figure*}

Tables~\ref{tab:0m} to~\ref{tab:5m} show the masses, in MeV, of the strange mesons predicted by our constituent quark model for spin-parity $J^P=0^-$ to $5^-$, respectively. Moreover, they compare our spectra with those reported by other theoretical approaches. Figure~\ref{fig:spectrum} shows a global picture of our kaon spectrum, comparing it with the states collected in PDG. If the kaon state is well-established and thus it appears in the PDG's summary table, the corresponding box is colored in blue, otherwise it is drawn in red. It is worth mentioning here that our spectrum is finite because the energy of the quark-antiquark bound-state cannot be larger than $m_q + m_{\bar q} + V_{\text{thr}} = 2591\,\text{MeV}$. There is only one kaon state listed in the PDG whose mass is larger than such a threshold. Particularly, the $K(3100)$, which was measured by only one collaboration in 1993~\cite{EXCHARM:1992fpf}, is omitted from the summary table of PDG and collected with the warning of needing confirmation.

We proceed now to describe in detail our theoretical findings. 

\begin{table}[!t]
\caption{\label{tab:0m} Masses, in MeV, predicted by our constituent quark model for kaon states with quantum numbers $J^P=0^-$. First column shows the name of the corresponding meson, second column refers to the spin-parity quantum numbers, third column is dedicated to the $n^{2S+1}L_J$ dominant partial wave in the kaon's wave function, fourth column shows our predicted mass and the successive columns collect the masses reported by PDG first and by other theoretical approaches later.}
\begin{ruledtabular}
\begin{tabular}{cccrr|rrr}
Meson & $J^{P}$ & Dom. & The. & Exp. & \cite{Pang:2017dlw} & \cite{Godfrey:1985xj} & \cite{Ebert:2009ub} \\
\hline
$K$ & $0^{-}$ & $1^1S_0$ & $481$ & $495$ & $498$ & $462$ & $482$ \\
& & $2^1S_0$ & $1512$ & $1482\pm16$ & $1457$ & $1454$ & $1538$ \\
& & $3^1S_0$ & $2018$ & $\cdots$ & $1924$ & $2065$ & $2065$ \\
& & $4^1S_0$ & $2318$ & $\cdots$ & $2248$ & $\cdots$ & $\cdots$ \\
& & $5^1S_0$ & $2488$ & $\cdots$ & $\cdots$ & $\cdots$ & $\cdots$ \\
& & $6^1S_0$ & $2567$ & $\cdots$ & $\cdots$ & $\cdots$ & $\cdots$ \\
\end{tabular}
\end{ruledtabular}
\end{table}

Besides the ground state kaon $K(495)$, the excited states with equal spin-parity $J^P=0^-$ collected by the PDG are $K(1460)$, $K(1630)$, $K(1830)$ and $K(3100)$. From those, only the first one is confirmed whereas the remaining three are omitted from the summary table, with $K(1630)$ and $K(3100)$ even having unknown quantum numbers. One can see in Table~\ref{tab:0m} that our constituent quark model predicts correctly the masses of the well-established $K(495)$ and $K(1460)$ mesons. The deviation with respect to the experimental values is less than $25\,\text{MeV}$, which is within the expected theoretical uncertainty. Moreover, our results for these two states are in reasonable agreement with those predicted by other theoretical frameworks, and this gives us confidence to contrast the rest of excited states. They also compare well and thus, based on our discussion, we do not believe that the observed $K(1630)$, $K(1830)$ and $K(3100)$ fit the quantum numbers $J^P=0^-$.

\begin{table}[!t]
\caption{\label{tab:0p} Masses, in MeV, predicted by our constituent quark model for kaon states with quantum numbers $J^P=0^+$. Note that we use the same column description as in Table~\ref{tab:0m}.}
\begin{ruledtabular}
\begin{tabular}{cccrr|rrr}
Meson & $J^{P}$ & Dom. & The. & Exp. & \cite{Pang:2017dlw} & \cite{Godfrey:1985xj} & \cite{Ebert:2009ub} \\
\hline
$K_0^\ast$ & $0^{+}$ & $1^3P_0$ & $1305$ & $1425\pm50$ & $1257$ & $1234$ & $1362$ \\
& & $2^3P_0$ & $1894$ & $1944\pm18$ & $1829$ & $1890$ & $1791$ \\
& & $3^3P_0$ & $2242$ & $\cdots$ & $2176$ & $2160$ & $2160$ \\
& & $4^3P_0$ & $2447$ & $\cdots$ & $2424$ & $\cdots$ & $\cdots$ \\
& & $5^3P_0$ & $2552$ & $\cdots$ & $\cdots$ & $\cdots$ & $\cdots$ \\
& & $6^3P_0$ & $2587$ & $\cdots$ & $\cdots$ & $\cdots$ & $\cdots$ \\ 
\end{tabular}
\end{ruledtabular}
\end{table}

The identification of scalar mesons is problematic because some of them have very large decay widths causing a strong overlap between resonances and background. Furthermore, in the mass range of interest, one should expect non-$q\bar q$ scalar systems such as glueballs, multiquark states and even kinematic phenomena like cusp effects due to the presence of close thresholds (the interested reader is referred to, for instance, Refs.~\cite{Close:2002zu, AMSLER200461, Bugg:2004xu, Klempt:2007cp, Pelaez:2015qba}). The $K_0^\ast(700)$, $K_0^\ast(1430)$ and $K_0^\ast(1950)$ are the three states collected in the PDG with scalar nature. All theoretical studies of $\pi K$ scattering that include constraints from chiral symmetry at low energies naturally find the $K _0^\ast(700)$ as a pole, with a mass less than $800\,\text{MeV}$ and a very large width of about $500\,\text{MeV}$. Its nature, however, does not seem to point only to a quark-antiquark bound-state, rather to a dynamically generated 4-quark resonance or a mixture of $2$- and $4$-quark structures. As other theoretical approaches, we have considered that our constituent quark model does not predict the $K _0^\ast(700)$ as a naive quark-antiquark bound state and thus the $K_0^\ast(1430)$ and $K_0^\ast(1950)$ mesons are assigned to the $1^3P_0$ and $2^3P_0$ states, respectively. Looking at Table~\ref{tab:0p}, our theoretical mass values are in reasonable agreement with those predicted by other approaches but suffer of discrepancies with respect to the experimental data; therefore, every formalism fails on describing quantitatively the scalar strange meson sector although ours seems to deliver a better agreement, but the nature of all $K_0^\ast$ candidates is still full of controversies.

\begin{table}[!t]
\caption{\label{tab:1p} Masses, in MeV, predicted by our constituent quark model for kaon states with quantum numbers $J^P=1^+$. Note that we use the same column description as in Table~\ref{tab:0m}.}
\begin{ruledtabular}
\begin{tabular}{cccrr|rrr}
Meson & $J^{P}$ & Dom. & The. & Exp. & \cite{Pang:2017dlw} & \cite{Godfrey:1985xj} & \cite{Ebert:2009ub} \\
\hline
$K_1$ & $1^{+}$ & $1^1P_1$ & $1370$ & $1253\pm7$ & $1364$ & $1352$ & $1294$ \\
& & $1^3P_1$ & $1455$ & $1403\pm7$ & $1377$ & $1366$ & $1412$ \\
& & $2^1P_1$ & $1925$ & $1650\pm50$ & $1840$ & $1897$ & $1757$ \\
& & $2^3P_1$ & $1971$ & $\cdots$ & $1861$ & $1928$ & $1893$ \\
& & $3^1P_1$ & $2260$ & $\cdots$ & $2177$ & $2164$ & $2164$ \\
& & $3^3P_1$ & $2286$ & $\cdots$ & $2192$ & $2200$ & $2200$ \\
& & $4^1P_1$ & $2458$ & $\cdots$ & $2422$ & $\cdots$ & $\cdots$ \\
& & $4^3P_1$ & $2471$ & $\cdots$ & $2434$ & $\cdots$ & $\cdots$ \\
& & $5^1P_1$ & $2556$ & $\cdots$ & $\cdots$ & $\cdots$ & $\cdots$ \\
& & $5^3P_1$ & $2561$ & $\cdots$ & $\cdots$ & $\cdots$ & $\cdots$ \\
& & $6^1P_1$ & $2589$ & $\cdots$ & $\cdots$ & $\cdots$ & $\cdots$ \\ 
& & $6^3P_1$ & $2590$ & $\cdots$ & $\cdots$ & $\cdots$ & $\cdots$ \\ 
\end{tabular}
\end{ruledtabular}
\end{table}

The mesons with quantum numbers $(I,J^P)=(1/2,1^+)$ that are summarized in the PDG are $K_1(1270)$, $K_1(1400)$ and $K_1(1650)$. Their experimental masses are somewhat lower than those predicted by our constituent quark model. However, the disagreement is alleviated as we advance in excited states taking into account that the so-called $K_1(1650)$ signal is observed experimentally as various peaks in $K^+\phi$ and $K \pi\pi$ invariant mass distributions in the $[1.6-1.9]\,\text{GeV}$ energy region. Concerning other theoretical tools, the predicted spectrum is better than ours in this channel; however, it should be remembered that our model parameters are completely constrained and the kaon spectrum is a pure prediction in our case, whereas the other theoretical approaches have performed a fit to the experimental data before providing the full spectrum. An example, it is the mixing between spin-singlet, $n^1P_1$, and spin-triplet, $n^3P_1$, states that is driven in our model by the so-called anti-symmetric spin-orbit interaction, given unequal quark and antiquark masses; and fixed in a phenomenological way by other approaches. The masses of the axial kaons are quite sensible to this mixing and we predict states with approximately 50\% of both $^1P_1$ and $^3P_1$ partial-waves in their wave function.

\begin{table}[!t]
\caption{\label{tab:1m} Masses, in MeV, predicted by our constituent quark model for kaon states with quantum numbers $J^P=1^-$. Note that we use the same column description as in Table~\ref{tab:0m}.}
\begin{ruledtabular}
\begin{tabular}{cccrr|rrr}
Meson & $J^{P}$ & Dom. & The. & Exp. & \cite{Pang:2017dlw} & \cite{Godfrey:1985xj} & \cite{Ebert:2009ub} \\
\hline
$K^\ast$ & $1^{-}$ & $1^3S_1$ & $900$ & $896$ & $896$ & $903$ & $897$ \\
& & $2^3S_1$ & $1676$ & $\cdots$ & $1548$ & $1579$ & $1675$ \\
& & $1^3D_1$ & $1787$ & $1718\pm18$ & $1766$ & $1776$ & $1699$ \\
& & $3^3S_1$ & $2112$ & $\cdots$ & $1983$ & $1950$ & $2156$ \\
& & $2^3D_1$ & $2173$ & $\cdots$ & $2127$ & $2251$ & $2063$ \\
& & $4^3S_1$ & $2372$ & $\cdots$ & $2287$ & $\cdots$ & $\cdots$ \\
& & $3^3D_1$ & $2408$ & $\cdots$ & $2385$ & $\cdots$ & $\cdots$ \\
& & $5^3S_1$ & $2516$ & $\cdots$ & $\cdots$ & $\cdots$ & $\cdots$ \\
& & $4^3D_1$ & $2537$ & $\cdots$ & $2573$ & $\cdots$ & $\cdots$ \\
& & $6^3S_1$ & $2576$ & $\cdots$ & $\cdots$ & $\cdots$ & $\cdots$ \\
& & $5^3D_1$ & $2585$ & $\cdots$ & $\cdots$ & $\cdots$ & $\cdots$ \\ 
\end{tabular}
\end{ruledtabular}
\end{table}

The strange mesons reported by PDG with spin-parity $J^P=1^-$ are $K^\ast(896)$, $K^\ast(1410)$ and $K^\ast(1680)$. One can see in Table~\ref{tab:1m} that there is agreement between theory and experiment for the masses of the $K^\ast(896)$ and $K^\ast(1680)$ mesons; being the last one traditionally assigned to the $1^3D_1$ state, but perfectly compatible with our prediction for the $2^3S_1$ case, and that reported in Ref.~\cite{Ebert:2009ub}. Note now that the mass of $K^\ast(1410)$ does not fit any theoretical result, with an experimental value $(150-300)\,\text{MeV}$ lower than the reported theoretical ones. Despite the PDG lists $K^\ast(1410)$ meson as a plausible candidate of the $2^3S_1$ state, one must face the following puzzles: (i) the mass of the $K^\ast(1410)$ is smaller than that of the experimentally assigned $2^1S_0$ state, and (ii) if $K^\ast(1410)$ is a member of the $2^3S_1$ nonet, together with $\rho(1450)$, $\omega(1420)$ and $\phi(1680)$, its mass is unexpectedly low from a theoretical point of view. All this makes us to be cautious with respect the nature of the $K^\ast(1410)$ meson and thus we leave it without a conventional quark-antiquark assignment.

\begin{table}[!t]
\caption{\label{tab:2m} Masses, in MeV, predicted by our constituent quark model for kaon states with quantum numbers $J^P=2^-$. Note that we use the same column description as in Table~\ref{tab:0m}.}
\begin{ruledtabular}
\begin{tabular}{cccrr|rrr}
Meson & $J^{P}$ & Dom. & The. & Exp. & \cite{Pang:2017dlw} & \cite{Godfrey:1985xj} & \cite{Ebert:2009ub} \\
\hline
$K_2$ & $2^-$ & $1^1D_2$ & $1760$ & $1773\pm8$ & $1778$ & $1791$ & $1709$ \\
& & $1^3D_2$ & $1854$ & $1819\pm12$ & $1789$ & $1804$ & $1824$ \\
& & $2^1D_2$ & $2160$ & $\cdots$ & $2121$ & $2238$ & $2066$ \\
& & $2^3D_2$ & $2214$ & $2247\pm17$ & $2131$ & $2254$ & $2163$ \\
& & $3^1D_2$ & $2403$ & $\cdots$ & $2380$ & $\cdots$ & $\cdots$ \\
& & $3^3D_2$ & $2432$ & $\cdots$ & $2388$ & $\cdots$ & $\cdots$ \\
& & $4^1D_2$ & $2535$ & $\cdots$ & $2570$ & $\cdots$ & $\cdots$ \\
& & $4^3D_2$ & $2547$ & $\cdots$ & $2575$ & $\cdots$ & $\cdots$ \\
& & $5^1D_2$ & $2585$ & $\cdots$ & $\cdots$ & $\cdots$ & $\cdots$ \\ 
& & $5^3D_2$ & $2588$ & $\cdots$ & $\cdots$ & $\cdots$ & $\cdots$ \\
\end{tabular}
\end{ruledtabular}
\end{table}

The PDG reports four candidates to be strange mesons with spin-parity $J^P=2^-$; they are $K_2(1580)$, $K_2(1770)$, $K_2(1820)$ and $K_2(2250)$. Among them, the first and last ones are omitted from the PDG's summary table  because distinct reasons. The $K_2(1580)$ meson was observed in $1979$ when performing a partial-wave analysis of the $K^-\pi^+\pi^-$ system, and needs confirmation. As one can see in Table~\ref{tab:2m}, none of the theoretical approaches locates the ground state in such mass region; in fact, all agree that the ground state should have a mass of around $1.75\,\text{GeV}$, which match the mass of the $K_2(1770)$ meson, making it the most plausible ground state of the strange mesons with $J^P=2^-$. The last mass measurement of the $K(1820)$ candidate was reported by the LHCb collaboration in 2017 with an experimental value of $(1853\pm27_{-35}^{+18})\,\text{MeV}$~\cite{LHCb:2016axx}, which coincides quite nicely with our model prediction; the PDG's average value is a bit smaller because a measurement of $1993$ with a mass value $1816\pm13$~\cite{ASTON1993186} is also taken into account. The $K_2(2250)$ signal contains various peaks in strange meson systems within the invariant mass energy region $2.15-2.26\,\text{GeV}$, as well as enhancements seen in the antihyperon-nucleon system, either in the mass spectra or in the $J^P=2^-$ wave. From a theoretical point of view, this multiplicity can be explained because the $2^1D_2$ and $2^3D_2$ states are close in mass, and within the energy range reported experimentally. Our constituent quark model predicts the following masses $2160\,\text{MeV}$ and $2214\,\text{MeV}$ for $2^1D_2$ and $2^3D_2$, respectively. They are also in the energy range predicted by other theoretical formalisms; in fact, the excited spectra are quite similar when comparison is possible.

\begin{table}[!t]
\caption{\label{tab:2p} Masses, in MeV, predicted by our constituent quark model for kaon states with quantum numbers $J^P=2^+$. Note that we use the same column description as in Table~\ref{tab:0m}.}
\begin{ruledtabular}
\begin{tabular}{cccrr|rrr}
Meson & $J^{P}$ & Dom. & The. & Exp. & \cite{Pang:2017dlw} & \cite{Godfrey:1985xj} & \cite{Ebert:2009ub} \\
\hline
$K_2^\ast$ & $2^+$ & $1^3P_2$ & $1454$ & $1427\pm2$ & $1431$ & $1428$ & $1424$ \\
& & $2^3P_2$ & $1975$ & $1994_{-50}^{+60}$ & $1870$ & $1938$ & $1896$ \\
& & $1^3F_2$ & $2095$ & $\cdots$ & $2093$ & $2151$ & $1964$ \\
& & $3^3P_2$ & $2290$ & $\cdots$ & $2198$ & $2206$ & $2206$ \\
& & $2^3F_2$ & $2364$ & $\cdots$ & $2356$ & $2551$ & $\cdots$ \\
& & $4^3P_2$ & $2474$ & $\cdots$ & $2438$ & $\cdots$ & $\cdots$ \\
& & $3^3F_2$ & $2519$ & $\cdots$ & $2560$ & $\cdots$ & $\cdots$ \\
& & $5^3P_2$ & $2563$ & $\cdots$ & $\cdots$ & $\cdots$ & $\cdots$ \\
& & $4^3F_2$ & $2584$ & $\cdots$ & $2695$ & $\cdots$ & $\cdots$ \\
\end{tabular}
\end{ruledtabular}
\end{table}

The $(I,J^P)=(1/2,2^+)$ channel is populated until now by two candidates, the $K_2^\ast(1430)$ and $K_2^\ast(1980)$ mesons reported by PDG. They have masses $(1427\pm2)\,\text{MeV}$ and $(1994_{-50}^{+60})\,\text{MeV}$, respectively, and match quite well our quark model predictions for the $1^3P_2$ and $2^3P_2$ states, as can be seen in Table~\ref{tab:2p}. It is worth to note here that the $K_2^\ast(1980)$ needs confirmation, but the agreement between the different theoretical approaches in the $J^P=2^+$ medium-energy spectrum makes us confident that $K_2^\ast(1980)$ is assigned correctly. As one can notice in Table~\ref{tab:2p}, the very high energy spectra differs between the two approaches that are able to deliver results. This is mostly due to the fact that the linear-screening confining interaction is fixed differently, in our case a global fitting from the light to the heavy quark sectors is pursued while in Ref.~\cite{Pang:2017dlw} only the kaon sector is considered.

\begin{table}[!t]
\caption{\label{tab:3p} Masses, in MeV, predicted by our constituent quark model for kaon states with quantum numbers $J^P=3^+$. Note that we use the same column description as in Table~\ref{tab:0m}.}
\begin{ruledtabular}
\begin{tabular}{cccrr|rrr}
Meson & $J^{P}$ & Dom. & The. & Exp. & \cite{Pang:2017dlw} & \cite{Godfrey:1985xj} & \cite{Ebert:2009ub} \\
\hline
$K_3$ & $3^+$ & $1^1F_3$ & $2047$ & $\cdots$ & $2075$ & $2131$ & $2009$ \\
& & $1^3F_3$ & $2132$ & $\cdots$ & $2084$ & $2143$ & $2080$ \\
& & $2^1F_3$ & $2340$ & $2324\pm24$ & $2340$ & $2524$ & $2348$ \\
& & $2^3F_3$ & $2387$ & $\cdots$ & $2347$ & $2536$ & $\cdots$ \\
& & $3^1F_3$ & $2507$ & $\cdots$ & $2550$ & $\cdots$ & $\cdots$ \\
& & $3^3F_3$ & $2529$ & $\cdots$ & $2546$ & $\cdots$ & $\cdots$ \\
& & $4^1F_3$ & $2581$ & $\cdots$ & $2688$ & $\cdots$ & $\cdots$ \\
& & $4^3F_3$ & $2587$ & $\cdots$ & $2691$ & $\cdots$ & $\cdots$ \\
\end{tabular}
\end{ruledtabular}
\end{table}

The strange mesons with spin-parity $J^P=3^+$ are characterized in its wave function by the presence of the $^1F_3$ and $^3F_3$ partial waves. Their mixture is still quite large, \emph{i.e.} it is around $50\%$, with a slightly dominance of one of them, exchanging as we go up in energy. Such a mixture for states with higher orbital angular momenta indicates that it is not a short-range interaction, producing pairs of states quite degenerate in mass even for $J=3$ (and higher), as can be verified in Table~\ref{tab:3p}. All theoretical computations predict ground states at $(2.05-2.14)\,\text{GeV}$ and first excited ones at $(2.32-2.54)\,\text{GeV}$, and the two cases able to provide very high excited states seem also to agree. The only meson collected in the PDG with quantum numbers $(I,J^P)=(1/2,3^+)$ is the $K_3(2320)$. It has been seen in the $J^P=3^+$ wave of the antihyperon-nucleon system but needs confirmation and thus it has been omitted from the summary table. In any case, the mass of the $K_3(2320)$ is $(2324\pm24)\,\text{MeV}$, which fits well the energy range corresponding to the first excited multiplet that contains the $2^1F_3$ and $2^3F_3$ states.

\begin{table}[!t]
\caption{\label{tab:3m} Masses, in MeV, predicted by our constituent quark model for kaon states with quantum numbers $J^P=3^-$. Note that we use the same column description as in Table~\ref{tab:0m}.}
\begin{ruledtabular}
\begin{tabular}{cccrr|rrr}
Meson & $J^{P}$ & Dom. & The. & Exp. & \cite{Pang:2017dlw} & \cite{Godfrey:1985xj} & \cite{Ebert:2009ub} \\
\hline
$K_3^\ast$ & $3^-$ & $1^3D_3$ & $1810$ & $1779\pm8$ & $1781$ & $1794$ & $1789$ \\
& & $2^3D_3$ & $2191$ & $\cdots$ & $2121$ & $2237$ & $2182$ \\
& & $1^3G_3$ & $2316$ & $\cdots$ & $2336$ & $2458$ & $2207$ \\
& & $3^3D_3$ & $2421$ & $\cdots$ & $2382$ & $\cdots$ & $\cdots$ \\
& & $2^3G_3$ & $2499$ & $\cdots$ & $2540$ & $2814$ & $\cdots$ \\
& & $4^3D_3$ & $2543$ & $\cdots$ & $2571$ & $\cdots$ & $\cdots$ \\ 
& & $3^3G_3$ & $2584$ & $\cdots$ & $2687$ & $\cdots$ & $\cdots$ \\
& & $5^3D_3$ & $2587$ & $\cdots$ & $\cdots$ & $\cdots$ & $\cdots$ \\ 
& & $4^3G_3$ & $\cdots$ & $\cdots$ & $2790$ & $\cdots$ & $\cdots$ \\
\end{tabular}
\end{ruledtabular}
\end{table}

With respect to the $J^P=3^+$ strange mesons, a complete different picture is found for the $(I,J^P)=(1/2,3^-)$ channel. Two partial waves also coexist in this case, the triplet-spin $D$ and $G$ waves, but its mixture is negligible, given by the tensor terms of the quark-antiquark potential. When comparison with other theoretical approaches is possible, our constituent quark model spectrum coincides quite well. For instance, considering all theoretical works, the ground state of the $(I,J^P)=(1/2,3^-)$ channel is predicted to have a mass between $1.78\,\text{GeV}$ and $1.81\,\text{GeV}$, see Table~\ref{tab:3m}. Moreover, the PDG provides a unique $J^P=3^-$ meson, $K_3^\ast(1780)$, whose mass $(1779\pm8)\,\text{MeV}$ makes it a perfect candidate to be the lowest-lying state. Analyzing the medium and high energy spectrum of our quark model and that predicted by Ref.~\cite{Pang:2017dlw}, \emph{i.e.} the most completed spectra, one can conclude that both are quite similar with more apparent discrepancies in the very high energy region.

\begin{table}[!t]
\caption{\label{tab:4m} Masses, in MeV, predicted by our constituent quark model for kaon states with quantum numbers $J^P=4^-$. Note that we use the same column description as in Table~\ref{tab:0m}.}
\begin{ruledtabular}
\begin{tabular}{cccrr|rrr}
Meson & $J^{P}$ & Dom. & The. & Exp. & \cite{Pang:2017dlw} & \cite{Godfrey:1985xj} & \cite{Ebert:2009ub} \\
\hline
$K_4$ & $4^-$ & $1^1G_4$ & $2270$ & $\cdots$ & $2309$ & $2422$ & $2255$ \\
& & $1^3G_4$ & $2337$ & $\cdots$ & $2317$ & $2433$ & $2285$ \\
& & $2^1G_4$ & $2476$ & $2490\pm20$ & $2520$ & $2779$ & $2575$ \\
& & $2^3G_4$ & $2510$ & $\cdots$ & $2526$ & $2789$ & $\cdots$ \\
& & $3^1G_4$ & $2576$ & $\cdots$ & $2673$ & $\cdots$ & $\cdots$ \\
& & $3^3G_4$ & $2587$ & $\cdots$ & $2677$ & $\cdots$ & $\cdots$ \\
& & $4^1G_4$ & $\cdots$ & $\cdots$ & $2782$ & $\cdots$ & $\cdots$ \\
& & $4^3G_4$ & $\cdots$ & $\cdots$ & $2785$ & $\cdots$ & $\cdots$ \\
\end{tabular}
\end{ruledtabular}
\end{table}

Concerning the $J^P=4^-$ channel, the $K_4(2500)$ state is the only one listed by the PDG but omitted from the summary table because it needs confirmation. Table~\ref{tab:4m} shows that the $K_4(2500)$ mass, $(2490\pm20)\,\text{MeV}$, is clearly higher than the one predicted by any theoretical approach for the ground state, either $1^1G_4$ or $1^3G_4$. Moreover, there is reasonable agreement between the different approaches with respect to the spectrum of the $(I,J^P)=(1/2,4^-)$ channel. Therefore, we assign the $K_4(2500)$ meson to the $2^1G_4$ state, without ruling out the possibility of being the $2^3G_4$ state, due to the fact that these two states are quite close in mass. Again, one can see that the spectrum predicted by Ref.~\cite{Pang:2017dlw} differs with ours at very large energies because the way of fitting the model parameters related with the linear-screening potential, \emph{i.e.} our kaon spectrum is a pure prediction where the model parameters have been fitted to the full meson spectra whereas that of Ref.~\cite{Pang:2017dlw} is obtained from a fit of some lowest-lying kaon states.

\begin{table}[!t]
\caption{\label{tab:4p} Masses, in MeV, predicted by our constituent quark model for kaon states with quantum numbers $J^P=4^+$. Note that we use the same column description as in Table~\ref{tab:0m}.}
\begin{ruledtabular}
\begin{tabular}{cccrr|rrr}
Meson & $J^{P}$ & Dom. & The. & Exp. & \cite{Pang:2017dlw} & \cite{Godfrey:1985xj} & \cite{Ebert:2009ub} \\
\hline
$K_4^\ast$ & $4^+$ & $1^3F_4$ & $2080$ & $2048_{-9}^{+8}$ & $2058$ & $2108$ & $2096$ \\
& & $2^3F_4$ & $2359$ & $\cdots$ & $2328$ & $2504$ & $2436$ \\
& & $1^3H_4$ & $2477$ & $\cdots$ & $\cdots$ & $\cdots$ & $\cdots$ \\
& & $3^3F_4$ & $2517$ & $\cdots$ & $2533$ & $\cdots$ & $\cdots$ \\
& & $2^3H_4$ & $2584$ & $\cdots$ & $\cdots$ & $\cdots$ & $\cdots$ \\
& & $4^3F_4$ & $2585$ & $\cdots$ & $2683$ & $\cdots$ & $\cdots$ \\
\end{tabular}
\end{ruledtabular}
\end{table}

Table~\ref{tab:4p} shows the $J^P=4^+$ strange meson spectrum predicted by our constituent quark model and compares it with those predicted by other theoretical approximations. There is good agreement for the ground-state mass, lying around $[2.06-2.11]\,\text{GeV}$. The $K_4^\ast(2045)$ is the only state collected in the PDG and we are pleased to see that its mass lies a little below the theoretical interval. There is agreement between our constituent quark model and Ref.~\cite{Pang:2017dlw} for the mass of the first excited state whereas Refs.~\cite{Godfrey:1985xj, Ebert:2009ub} predict such state 100-150 MeV higher. Concerning the medium and high energy mass range, our model and that of Ref.~\cite{Pang:2017dlw} provide similar spectra except for the latest state, probably due to the different way of fitting the confining interaction at large inter-quark distances.

\begin{table}[!t]
\caption{\label{tab:5p} Masses, in MeV, predicted by our constituent quark model for kaon states with quantum numbers $J^P=5^+$. Note that we use the same column description as in Table~\ref{tab:0m}.}
\begin{ruledtabular}
\begin{tabular}{cccrr|rrr}
Meson & $J^{P}$ & Dom. & The. & Exp. & \cite{Pang:2017dlw} & \cite{Godfrey:1985xj} & \cite{Ebert:2009ub} \\
\hline
$K_5$ & $5^+$ & $1^1H_5$ & $2442$ & $\cdots$ & $\cdots$ & $\cdots$ & $\cdots$ \\
& & $1^3H_5$ & $2489$ & $\cdots$ & $\cdots$ & $\cdots$ & $\cdots$ \\
& & $2^1H_5$ & $2572$ & $\cdots$ & $\cdots$ & $\cdots$ & $\cdots$ \\
& & $2^3H_5$ & $2589$ & $\cdots$ & $\cdots$ & $\cdots$ & $\cdots$ \\
\end{tabular}
\end{ruledtabular}
\end{table}

Unfortunately, there is no experimental data neither theoretical computations to compare our spectrum of strange mesons with spin-parity $J^P=5^+$. This is mostly due to the fact that these states have very high orbital angular momentum with either spin zero or one. As shown in Table~\ref{tab:5p}, we predict four states, which are very close in mass, until reaching threshold. The ground-state mass is $2442\,\text{MeV}$ and the remaining three excited states have masses of $2489\,\text{MeV}$, $2572\,\text{MeV}$ and $2589\,\text{MeV}$, respectively. That is to say, there is only $150\,\text{MeV}$ between the lowest and highest masses.

\begin{table}[!t]
\caption{\label{tab:5m} Masses, in MeV, predicted by our constituent quark model for kaon states with quantum numbers $J^P=5^-$. Note that we use the same column description as in Table~\ref{tab:0m}.}
\begin{ruledtabular}
\begin{tabular}{cccrr|rrr}
Meson & $J^{P}$ & Dom. & The. & Exp. & \cite{Pang:2017dlw} & \cite{Godfrey:1985xj} & \cite{Ebert:2009ub} \\
\hline
$K_5^\ast$ & $5^-$ & $1^3G_5$ & $2291$ & $2382\pm24$ & $2286$ & $2388$ & $2356$ \\
& & $2^3G_5$ & $2488$ & $\cdots$ & $2504$ & $2749$ & $2656$ \\
& & $1^3I_5$ & $2581$ & $\cdots$ & $\cdots$ & $\cdots$ & $\cdots$ \\
& & $3^3G_5$ & $2588$ & $\cdots$ & $2662$ & $\cdots$ & $\cdots$ \\
& & $2^3I_5$ & $\cdots$ & $\cdots$ & $\cdots$ & $\cdots$ & $\cdots$ \\
& & $4^3G_5$ & $\cdots$ & $\cdots$ & $2776$ & $\cdots$ & $\cdots$ \\
\end{tabular}
\end{ruledtabular}
\end{table}

The $K_5^\ast(2380)$ is the only state reported by PDG with quantum numbers $(I,J^P)=(1/2,5^+)$. It was discovered in the $1986$ and not confirmed by any other experimental collaboration. This is the reason because it has been omitted from the summary table. Its mass is $(2382\pm24)\,\text{MeV}$ lies just in the middle of the masses predicted by our constituent quark model for the $1^3G_5$ and $2^3G_5$ states. The same situation is found in the latest work related with kaon spectrum~\cite{Pang:2017dlw}; however, it coincides in mass with the ground states reported by Refs.~\cite{Godfrey:1985xj, Ebert:2009ub}. This feature could be related with the slope of the linear confining potential but it is difficult to make a strong statement without having more data related with this state.

Finally, we show the complete kaon spectrum predicted by our constituent quark model in Figure~\ref{fig:spectrum}. It also compares the different states with those listed in the PDG with the following legend of colors: the states collected in the summary table are colored in blue whereas those listed by the PDG but omitted from the summary table are shown in red. The states are grouped by their spin-parity $J^P$ quantum numbers, and named accordingly. The dashed line in the figure represents the energy threshold of quark-antiquark bound-states. It is clearly shown in the figure, and common to any theoretical approach, that the kaon spectrum is rich with many quark model states not yet seen experimentally. Moreover, exotic states in this sector are expected and must be added to this quark-antiquark picture. From an experimental point of view, there are $25$ strange mesons that have been detected; however, only $14$ are well established and included in the summary tables whereas $11$ states still need further confirmation. Among the $14$ well-established kaons, our quark model picture faces fitting problems with only two of them: $K^\ast(1410)$ and $K_1(1650)$, which usually present issues to fit in other theoretical frameworks and they are candidates for exotic states. Only three of the eleven states that are omitted from the summary table do not present natural assignments within our quark model. They are $K_0^\ast(700)$, $K(1630)$ and $K_2(1580)$. The first one shows a nature not compatible with a quark-antiquark state, the second one suffers from even not fixed its spin-parity quantum numbers, and the third one is an isolated measure performed in 1979.


\section{Summary}
\label{sec:summary}

A completely new horizon in kaon physics has been opened with the CERN's approval of upgrading the external M2 beam line of the CERN SPS in order to provide radio-frequency separated high-intensity and high-energy kaon and antiproton beams. The kaon beam, in combination with a universal spectrometer of the COMPASS experiment, will allow to map out the complete spectrum of excited kaons with unprecedented precision, using novel analysis methods. 

In support of the experimental effort, we have revisited herein the kaon spectrum using a constituent quark model in which the quark-antiquark interaction (based on the non-perturbative phenomena of dynamical chiral symmetry breaking and color confinement, plus the perturbative one-gluon exchange force) has been fitted elsewhere through hadron, hadron-hadron and multiquark phenomenology, from light to heavy quark sectors, and thus the model parameters are completely constrained. The bound-state problem is solved using the Rayleigh-Ritz variational method in which the genuine state's wave function is expanded by means of Gaussian trial functions whose ranges are in geometric progression.

We have compare our results with those available in the literature in order to provide some general statements, common to all calculations. In addition, we have faced our spectrum to the states listed by the PDG. Among the results we have described, the following are of particular interest:
First, it is common to any theoretical approach that the kaon spectrum is rich with many quark model states not yet seen experimentally.
Second, our quark model picture faces fitting problems with only two of the fourteen well-established kaons. The $K^\ast(1410)$ and $K_1(1650)$ are usually difficult to describe by any theoretical computation and thus they are candidates for exotic states.
Third, only three of the eleven states that are omitted from the summary table do not present natural assignments within our quark model. They are $K_0^\ast(700)$, $K(1630)$ and $K_2(1580)$. The first one shows a nature not compatible with a quark-antiquark state, the second one suffers from even not fixed its spin-parity quantum numbers, and the third one is an isolated measure performed in 1979.
And fourth, when comparison with other theoretical approaches is possible, most of the states in the low- and medium-energy region compare nicely, whereas the (very) high-energy region is characterized differently because assumed confining interaction.

It is important to remark that our predicted \emph{na\"ive} quark-antiquark bound states can be influenced by nearby meson-meson thresholds and thus their composition, binding energy and decay properties can be modified. The expansion of the Fock’s space including, together with the quark-antiquark degrees of freedom, states of four (anti-)quarks has been done within our approach in the heavy quark sectors and it is expected to carry out kindred studies in the kaon spectrum, where such effects appear to be even more important, because the ubiquitous Goldstone-boson exchange interactions, and could explain the discrepancies found herein between theory and experiment. This task, however, is beyond the scope of the present manuscript.


\begin{acknowledgments}
This work has been partially funded by
EU Horizon 2020 research and innovation program, STRONG-2020 project, under grant agreement no. 824093;
Ministerio Espa\~nol de Ciencia e Innovaci\'on, grant no. PID2019-107844GB-C22 and PID2019-105439GB-C22;
and Junta de Andaluc\'ia, contract nos. P18-FR-5057 and Operativo FEDER Andaluc\'ia 2014-2020 UHU-1264517.
\end{acknowledgments}


\bibliography{PrintKaons}

\end{document}